\documentclass[amsmath,amssymb,nofootinbib,prl,reprint]{revtex4-1}
\usepackage{dcolumn}
\usepackage{bm}

\usepackage{hyperref}
\usepackage{latexsym}
\usepackage[dvips]{graphicx}
\usepackage{epsf}
\usepackage{color}

\allowdisplaybreaks 
\newcommand{\be}{\begin{equation}}
\newcommand{\ee}{\end{equation}}
\newcommand{\bea}{\begin{eqnarray}}
\newcommand{\eea}{\end{eqnarray}}

\newcommand{\f}{\frac}

\let\a=\alpha     
     \let\th=\theta   \let\l=\lambda
\let\m=\mu    \let\n=\nu         \let\r=\rho   
\let\s=\sigma      
    
\let\G=\Gamma \let\D=\Delta   \let\L=\Lambda 
         
  \let\eps=\epsilon

\let\==\equiv

\newcommand{\na}{\nabla}
\newcommand{\Tr}{{\rm Tr}}

\newcommand{\hb}{\bar{h}}

\newcommand{\ft}{\tilde{f}}

\newcommand{\Rt}{\tilde{R}}

\newcommand{\cR}{\mathcal{R}}

\newcommand{\cT}{\mathcal{T}}

\begin{document}
%------------------------------------------------------------------------------

\begin{flushright} \small
AEI-2013-032
\end{flushright}

\title{{\bf On the number of relevant operators in asymptotically safe gravity}}
\author{Dario Benedetti}
%\email{dario.benedetti@aei.mpg.de}
\affiliation{ {\footnotesize Max Planck Institute for Gravitational Physics (Albert Einstein Institute), \\
Am M\"{u}hlenberg 1, D-14476 Golm, Germany }}

\begin{abstract}
The asymptotic safety scenario of gravity conjectures that ({\it i}) the quantum field theory of gravity exists thanks to the presence of a non-trivial ultraviolet fixed point of the renormalization group, and that  ({\it ii}) the fixed point has only a finite number of relevant perturbations, i.e. a finite number of UV-stable directions (or in other words, a finite number of free parameters to be fixed experimentally).
Within the $f(R)$ approximation of the functional renormalization group equation of gravity, we show that assuming the first half of the conjecture to be true, the remaining half follows from general arguments, that is, we show that assuming the existence of a non-trivial fixed point, the fact that the number of relevant directions is finite is a general consequence of the structure of the equations.
\end{abstract}

%\pacs{}

\maketitle

%------------------------------------------------------------------------------

%------------------------------------------------------------------------------
%{\it Introduction} --

The main problem with the perturbative non-renormalizability of gravity is notoriously the proliferation of couplings to be determined by experiments, a situation that severely limits the predictive power of perturbation theory. From a renormalization group point of view, this is understood as the fact that Newton's constant is technically an irrelevant coupling (i.e. it is on a UV-unstable trajectory) for the free (Gaussian) fixed point of Einstein's theory, and if we want to keep it finite in the continuum limit, we have to deal also with the infinitely many other irrelevant couplings. One solution to this problem was suggested long ago by Weinberg \cite{Weinberg:1976xy,Weinberg:1980gg}, who dubbed it {\it asymptotic safety}: our near-Gaussian unstable trajectory could be the UV-stable trajectory of a new non-Gaussian fixed point (NGFP). In order to be effective, such scenario requires that  ({\it i}) there exists a NGFP, and  that ({\it ii}) the number of parameters needed to uniquely determine one such trajectory among all the possible ones be finite. We associate such parameters to {\it relevant directions}, i.e. to a basis of independent trajectories spanning the UV-stable surface of the NGFP. If the dimension of the UV-stable surface was infinite, we would of course be confronted again with a problem similar to the one we started from. On the other hand, if it was finite, we would have the possibility of constructing a nonperturbatively renormalizable quantum field theory of gravity
with full predictive power.

An important amount of evidence has been collected in recent years in favor of the asymptotic safety scenario, mainly by studying truncations of a functional renormalization group equation (FRGE) \cite{Niedermaier:2006wt,Percacci:2007sz,Litim:2011cp,Reuter:2012id}. The adopted strategy (avoiding a perturbative expansion in the couplings) is to truncate the infinite-dimensional theory space of all possible effective actions to a finite-dimensional subspace, to look for fixed points and their relevant directions, and eventually, after subsequently increasing the truncation and repeating the procedure, to look for evidence of convergence.
Such program has been carried out to a certain extent, in particular with truncations of the effective Lagrangian to a polynomial in the Ricci scalar $R$, up to order $R^8$ \cite{Lauscher:2002sq,Codello:2007bd,Machado:2007ea,Codello:2008vh}, and more recently up to order $R^{34}$ \cite{Falls:2013new}, resulting always in a fixed point with only three relevant directions.
However, lacking a more general understanding of such empirical observations, it has up to now remained an open question whether in the full theory the number of relevant directions would remain finite, and hence whether the predictive power of the scenario would survive.
It is the purpose of this Letter to address this point, and to show,  within the context of an $f(R)$ approximation, that a proof can be given for the finiteness of the UV-stable surface at any given fixed point of gravity.

In order to be able to make general statements about higher orders in a truncation to polynomials in $R$, it is essential to make one step further, and study a truncation of the theory space to an infinite dimensional subspace, described by a generic $f(R)$ Lagrangian, an approximation that was suggested in \cite{Benedetti:2012dx} as an analogue of the local potential approximation in scalar field theory (see \cite{Morris:1998da,Bagnuls:2000ae,Delamotte-review}).
The program of investigating the asymptotic safety scenario in such an approximation is still at the beginning, however it has already showed its power when it comes to disposing of spurious fixed points and in the understanding of the general characteristics of the fixed-point theory \cite{Benedetti:2012dx,Demmel:2012ub,Dietz:2012ic}.
Here, we will see that in addition it allows us to answer the question about the number of relevant directions in general terms.

%------------------------------------------------------------------------------
%{\it Body} --

The FRGE reads (here $\G_k$ is the effective action at scale $k$, and $t=\ln (k/\L)$, with $\L$ an initial scale)
\be\label{FRGE}
\f{d}{dt} \G_k = \f12 {\rm STr} \left[ \left( \G_k^{(2)} 
+ \cR_k \right)^{-1} \, \f{d}{dt}  \cR_k  \right] \, .
\ee
For its derivation, meaning and usage, we refer to the general reviews \cite{Niedermaier:2006wt,Percacci:2007sz,Litim:2011cp,Reuter:2012id,Morris:1998da,Bagnuls:2000ae,Delamotte-review,Gies:2006wv}
Here we emphasize only some aspects which are important for our work.
$ \cR_k $ is an IR cutoff operator defining the coarse graining scheme. The FRGE clearly depends on the choice of such scheme, however a number of universal properties of the flow should be independent of it, in particular the critical exponents, and hence the number of relevant directions at a fixed point.
Unfortunately, approximations spoil universality to some extent, and one has to be careful in analyzing different schemes in order to pinpoint eventual artifacts of particular schemes.
Scheme dependence can also be used to our advantage, optimizing the convergence of approximations to the exact results  \cite{Litim:2001up}.
In any case, a good cutoff should ensure that $\G^{(2)}+\cR_k$ be invertible. More precisely, being the second variation of a Legendre transform, it should be positive (remember that $\G_k$  on its own is not a Legendre transform, thus it needs not be convex), and have a gap at finite $k$ \cite{Litim:2001up}. 

Our approximation consists in projecting the FRGE for gravity on a maximally symmetric background, in particular on a four-dimensional sphere.\footnote{Everything goes through identically on the hyperboloid, apart from the absence of zero modes and the explicit appearance of volume factors, which anyway cancel out once the traces are evaluated.} As a result, any action terms depending on the Weyl tensor, on the traceless Ricci tensor, or on derivatives of the Ricci scalar, vanish identically, and we will only be able to study the running of an $f(R)$ theory, which hence we take as our ansatz for the effective action:  $\G_k= \int d^4x \sqrt{g} f_k(R)$ (plus gauge-fixing and ghosts \cite{Benedetti:2012dx}). 
We will not make any further approximation, and we will not truncate the Lagrangian to a polynomial in $R$.

For technical convenience, in gravity the common scheme is to adapt the cutoff to the truncation by taking $\cR_k(\D)=\G^{(2)}(P_k)-\G^{(2)}(\D)$, where
$P_k\equiv \D +k^2\, r(\D/k^2)$, $\D$ is a Laplace-type operator appearing in the second variation of the action (at least when gauge-fixing, field decompositions and background choice allow us to reduce all the differential operators to Laplace type), and $r(z)$ is a cutoff profile function. Such {\it adaptive cutoff} brings many advantages in the evaluation of the functional traces, however it also leads to a number of complications from the point of view of the resulting differential equation for $f(R)$.
In order to avoid such complications, 
we will differ here from previous works on the $f(R)$ approximation in the choice of cutoff scheme, by taking  an $f(R)$-independent cutoff.
Our choice has a crucial consequence: the resulting fixed-point differential equation will be of second order (as explained in \cite{Benedetti:2012dx}, the equations derived so far were of third order precisely because of cutoff choices with an $f(R)$ dependence).

We adopt the same notation and construction as in  \cite{Benedetti:2012dx}, where the reader can find all the details omitted here 
(field components, functional variations, gauge-fixing, etc.), only differing for the choice of absorbing Newton's constant inside $f(R)$, and for the cutoff scheme.
Defining the operators  $\D_0 \equiv -\na^2 - R/3$, $\D_1 \equiv -\na^2 - R/4$, and $\D_2 \equiv -\na^2 + R/6$, for the scalar, vector and tensor modes, respectively,
the fixed-point FRGE in the $f(R)$ approximation reads (as usual $f'=df/dR$, etc.)
\be \label{evalFRGE}
\tilde{V} \left(4 \ft_k(\Rt) - 2 \Rt \ft_k'(\Rt)    \right) = \cT_2 + \cT^{\hb}_0 + \cT^{\text{Jac}}_1 +\cT^{\text{Jac}}_0 \, ,
\ee
where tildes stand for dimensionless quantities, in particular $\ft_k(\Rt) = k^{-4} f_k(k^2 \Rt)$ and $\tilde{V}=k^4\int d^4x\sqrt{g}$ (on the sphere $\tilde{V}=384\pi^2/\Rt^2$), and
we have subdivided the rhs into the contributions of the transverse-traceless tensor modes (we define $E(R)=2 f(R)-R f'(R)$, which is zero on shell)
\be
\cT_2 = \Tr\left[ \frac{\f{d}{dt} \cR_k^{T} (\D_2+\a_2 R)}{ - f'(R) \D_2- E(R)/2+ 2 \cR_k^{T}(\D_2+\a_2 R)} \right] \, ,
\ee
the gauge-invariant trace mode $\hb$
\begin{widetext}
\be
\begin{split}
& \cT^{\hb}_0 = \Tr\Big[  \left. \frac{8 \f{d}{dt}\cR_k^{\hb} (\D_0+\a_0 R)  }{ 9 f''(R) \D_0^2+3 f'(R) \D_0 +E(R)+16  \cR_k^{\hb} (\D_0+\a_0 R) } \right] \, ,
\end{split}
\ee
\end{widetext}
the transverse vector modes
\be
\cT^{\text{Jac}}_1 = -\f12 \Tr\left[ \frac{ \f{d}{dt}\cR_k^{V} (\D_1+\a_1 R)}{\D_1+ \cR_k^{V} (\D_1+\a_1 R)} \right] \, ,
\ee
and the remaining scalar modes 
\be
\begin{split}
\cT^{\text{Jac}}_0 = & \f12 \Tr\left[ \frac{ \f{d}{dt}\cR_k^{S_1} (\D_0+\a_0 R) }{ \D_0+\frac{R}{3}+\cR_k^{S_1} (\D_0+\a_0 R)} \right] \\
 & -\Tr\left[  \frac{ 2 \f{d}{dt}\cR_k^{S_2} (\D_0+\a_0 R) }{ \left( 3 \D_0+R\right)
   \D_0+ 4 \cR_k^{S_2} (\D_0+\a_0 R)} \right] \, ,
\end{split}
\ee
the last two contributions both arising from the Jacobians of the field decompositions (longitudinal and ghosts contributions cancel out \cite{Benedetti:2012dx}).
Note that the traces are dimensionless despite being written in terms of dimensionful variables.
A crucial observation is that in order to properly implement the cutoff on all the modes, we should choose the $\a_s$ parameters such that $\D_s+\a_s R>0$ for all the modes, $s=0,1,2$ (we remind that on the sphere $\D_s\propto R$). 
In \cite{Benedetti:2012dx} we had chosen $\a_s=0$ in order to avoid certain poles that appeared in previous equations.
However, the appearance of such poles is associated to the adaptive cutoff scheme. We noticed in \cite{Benedetti:2012dx} that, within such scheme, in order to avoid poles in the rhs of \eqref{FRGE}, modes of rank $s$ had to be cut off with respect to the eigenvalues of $\D_s$.
In the present work we employ a different scheme, and there will be no concern about such singularities. For tensor and vector modes it is safe to take $\a_2=0$ and $\a_1=0$, as the spectra of $\D_2$ and $\D_1$ are strictly positive, while we have to be more careful with the scalar modes: on the sphere, the trace in the $\hb$ sector includes a constant mode $\hb^{(0)}$ for which $\D_0 \hb^{(0)} = -\f{R}{3} \hb^{(0)}$, hence we need to take $\a_0>1/3$.\footnote{On the hyperboloid we get the additional constraint $\a_0<25/48$.}
As already mentioned, universal properties of the RG flow should not depend on the cutoff scheme, however a poor choice of cutoff can lead to poor results. As observed in \cite{Benedetti:2012dx}, in the adaptive cutoff scheme, a fixed singularity at $\Rt=0$ renders the third-order equation effectively second order, thus pointing in favor of scheme independence, while fixed singularities at other locations are due to not properly imposing the cutoff on the lowest scalar modes.

We now choose the simple cutoff form $\cR_k^{\phi}(\D_s) = k^{m_{\phi}} c_\phi r(\tilde{\D}_s+\a_s \Rt)$, where $\phi$ labels a rank-$s$ field to which the cutoff is associated, $r(z)$ is a dimensionless profile function, identical for all the fields, $c_\phi$ is a (positive) free parameter, and the power $m_\phi$ is to be chosen so that the cutoff has the same dimension as the Hessian to which it is associated. The profile function should satisfy some basic requirements that make it a good IR cutoff, 
in particular it should be non-negative, monotonically decreasing and it should satisfy $\lim_{z\to 0}r(z)>0$ and $\lim_{z\to\infty}r(z)=0$. Common choices of profile functions are $r(z)=z\, (\exp(a z^b)-1)^{-1}$ (with $a>0$, $b\geq1$) \cite{Wetterich:1992yh}, or $r(z) = (1-z) \th(1-z)$ \cite{Litim:2001up}, but many more are of course possible. We will exclude power-law profile functions \cite{Morris:1994ie}, and we will assume that the approach to zero for $z\to\infty$ is faster than any power (power-law profile functions could however be used taking care of choosing a sufficiently high power). Special care should also be taken for non-analytic cutoffs (e.g. with step functions), and for simplicity we will assume strictly positive analytic profile functions.

For our purpose, it will be sufficient to study here only the large-$\Rt$ properties of the FRGE, for which we will not need to actually choose a specific cutoff profile and to perform the traces.
In this respect, one should notice that unlike in other applications of the FRGE, in the case of gravity there is a field dependence also in the operator with respect to which modes are being cut off (this aspect has been highlighted in a simple setting in \cite{Reuter:2008wj}), in particular $\D_s\propto R$ on the sphere, hence the large-field limit is peculiarly intertwined to the large mode suppression.

In analyzing the asymptotic behavior of the NGFP solution, we will assume that this is power-law. A justification comes both from experience and from physical considerations, as only to such behavior we can associate a familiar interpretation in terms of couplings \cite{Morris:1998da,Morris:1996xq}. 

Given such assumption on the asymptotics of the solution, we can study the dominant balancing of terms in the FRGE in the asymptotic regime. We find that the lhs of \eqref{evalFRGE}, as well the cutoff-independent parts of the denominators on its rhs, contribute in the large $\Rt$ limit with a power-law behavior.
On the other hand, the presence of the cutoff implies a faster fall-off of the rhs at large $\Rt$. 
As a consequence, the leading asymptotic behavior of the solution is dictated only by the lhs,
and at leading order the large-$\Rt$ equation reduces to
\be \label{evalFRGE_0}
4 \ft_k(\Rt) - 2 \Rt \ft_k'(\Rt)   = 0 \, ,
\ee
corresponding to classical scale invariance of the action.
Its solution is $\ft^*(\Rt)\sim \Rt^2$, and we recover the leading order of the asymptotic expansion found in \cite{Benedetti:2012dx,Dietz:2012ic}.
We would need to study the full equation, not just its asymptotics, in order to determine whether a global solution with such asymptotic behavior exists. 
We leave this problem to future work (for preliminary studies in alternative schemes see \cite{Benedetti:2012dx,Dietz:2012ic}), and take the existence of such a global solution as our main assumption here.

Next, we use the asymptotic behavior of the FP solution in order to study the equation for the linear perturbations in the large-$\Rt$ limit.
Linearization in the neighborhood of the fixed point is performed by writing
\be \label{linpert}
\ft_k(\Rt) \sim \ft^*(\Rt) + \eps\, v(\Rt) e^{-\th t} \, ,
\ee
and expanding the FRGE to linear order in $\eps$. The zeroth order is identically zero by construction, while the first order provides the equation for the perturbations,
which takes the form of an eigenvalue equation ($\l\equiv 4-\th$):
\be \label{eigEq}
 - a_2(\Rt) v''(\Rt) + a_1(\Rt)  v'(\Rt) + a_0(\Rt)  v(\Rt)  =   \l \, v(\Rt)  \,  .   
\ee
In the large-$\Rt$ limit, $a_0$ and $a_2$ go to zero faster than power-law, while $a_1\sim2\Rt$, and
as a consequence at leading order  $v(\Rt)\sim \Rt^{2-\th/2}$ for perturbations with  power-law asymptotics.

Perturbations with ${\rm Re}(\th)>0$ correspond to relevant directions, hence we want to prove that there is a finite number of eigenfunctions with $\l<4$.
We will actually show that the eigenvalues $\l$ form a real and discrete spectrum, bounded from below, and with a finite number of eigenfunctions with positive $\th$.
In order to accomplish that, we need only few more general properties of the coefficients $a_0$, $a_1$ and $a_2$.

\newblock
First we note that the coefficients have no singularities, a direct consequence of the assumption that a global solution $\ft^*(\Rt)$ exists, and of the presence of the IR cutoff in the FRGE.
Second, we observe that, due to the positivity ($r(z)>0$) and monotonicity ($r'(z)=dr(z)/dz<0$) of the cutoff, 
\begin{widetext}
\be \label{a2}
\begin{split}
& a_2 = \f{144 c_{\hb} }{\tilde{V}} \,  \Tr\Big[   \left. \frac{\tilde{\D}_0^2\, (2\, r (\tilde{\D}_0+\a_0 \Rt)- (\tilde{\D}_0+\a_0 \Rt)\, r' (\tilde{\D}_0+\a_0 \Rt) ) }{ (9 \ft''(\Rt) \tilde{\D}_0^2+3 \ft'(\Rt) \tilde{\D}_0+\tilde{E}(\Rt)+16  c_{\hb} r (\tilde{\D}_0+\a_0 \Rt))^2 } \right]  >0 \, .
\end{split}
\ee
\end{widetext}
Hence \eqref{eigEq} is a Sturm-Liouville problem, written with the usual sign convention, and with no singularities. 
The boundary conditions are provided by the requirement that the asymptotic behavior be power-law at $\Rt\sim\pm\infty$. These boundary conditions are equivalent to requiring square integrable solutions of \eqref{eigEq} with respect to the  weight function $w(\Rt)= a_2^{-1} \exp( -\int^{\Rt} \f{a_1}{a_2} )$, and they ensure that the Sturm-Liouville operator is self-adjoint, and hence that its spectrum is real.

In order to prove the existence of a discrete spectrum we can transform \eqref{eigEq} to a standard Schr\"odinger eigenvalue equation $-d^2 y(x)/dx^2 +U(x) y(x) = \l\, y(x)$, by means of a Liouville transform, and then apply standard theorems (e.g. \cite{Berezin-book}).
Defining the new variable $x=\int^{\Rt} 1/\sqrt{a_2}$ (with $\int^{\pm\infty} 1/\sqrt{a_2} =\pm\infty$), and substituting $y = a_2^{1/4} w^{1/2} v$,
we find the potential
\be \label{potential}
U(x) = a_0 + \f{a_1^2}{4 a_2} -\f{a_1'}{2} + a_2' \left(\f{a_1}{2a_2}+\f{3 a_2'}{16 a_2} \right) - \f{a_2''}{4} \, .
\ee
The potential has no singularities at finite $x$, as a consequence of \eqref{a2} and of the absence of singularities in the original equation.
Finally, the asymptotic behavior of  $a_0$, $a_1$ and $a_2$ is such that for $x\to\pm\infty$ the second term dominates, and $U(x)\to+\infty$.
These simple observations imply that the spectrum is discrete, bounded from below, and the only accumulation point is at infinity  \cite{Berezin-book}.
As a consequence, there is a finite number of eigen-perturbations with $\th>0$.

We have reached our goal of showing that, assuming the existence of a fixed point solution $\ft^*(\Rt)$, the number of relevant directions is finite, thus lending theoretical understanding to the empirical observation that the their number does not seem to grow with the order of the truncation in the polynomial case \cite{Codello:2007bd,Machado:2007ea,Codello:2008vh,Falls:2013new}.
Importantly, we found here that the exponents $\th$ are all real, contrary to what observed in polynomial truncations,
but compatibly with what observed in \cite{Demmel:2012ub,Dietz:2012ic} and in \cite{BMS1,BMS2}, and we conclude that complex exponents are probably an artifact of the truncations.

%------------------------------------------------------------------------------
%{\it Conclusions} -- 

We close with some general remarks.
Studying the limit $\Rt\to\infty$ of the fixed point solution, as explained in  \cite{Benedetti:2012dx}, means studying the limit $k\to 0$ at fixed $R$ (see also \cite{Delamotte-review} for a clear explanation of this aspect in the scalar case). 
As argued in \cite{Benedetti:2012dx}, the asymptotic behavior   $\ft^*(\Rt)\sim \Rt^2$ of the fixed point solution implies that the full effective action (obtained for $k\to 0$, i.e. with all the modes integrated out) at the fixed point is the scale invariant theory defined by $\G^*=\G^*_{k=0}=A^* \int d^4 x\sqrt{g}\, R^2$, with the constant $A^*$ to be determined by the requirement that $\G^*_k$ be non singular at all $\Rt$ (or at all $k$). Note that this expression is valid only on a maximally symmetric space, hence it should be interpreted with care: if we expect the fixed point to have conformal (or Weyl) invariance, then the only local Lagrangian satisfying such criterion, and reducing to $R^2$ in case of maximal symmetry, is given by the Gauss-Bonnet term, corresponding to a purely topological theory.\footnote{Conversations with R. Percacci on this observation are gratefully acknowledged.} While this might stimulate some speculations on the possibility of a topological fixed point, one should refrain from attaching much interpretation along these lines in our case as in the $f(R)$ approximation we are of course not seeing other possible terms like the Weyl-squared one, $C_{\m\n\r\s}C^{\m\n\r\s}$,
whose effects have up to now only been studied in finite truncations \cite{BMS1,BMS2}.

Going back to \eqref{linpert}, an infinitesimal $\eps$ ensures that at $t=0$, i.e. at the initial scale $k=\L$, $\ft_k(\Rt)$ is very close to the fixed point solution.
Integrating towards $k=0$, and discarding deviations from the linearized flow, we obtain the effective action
\be
\G_k \to \int d^4x \sqrt{g}  \{ A R^2 + \sum_i \eps_i\, \L^{\th_i} R^{2-\th_i/2}  \} \, .
\ee
In order to take $\L\to\infty$ while keeping the action finite, in the case of positive $\th$, we need to take $\eps\sim (m_\th/\L)^\th$, for some finite mass parameter $m_\th$.
For negative $\th$, the perturbations are automatically small in the large-$\L$ limit, without any fine tuning, i.e. they are irrelevant. Finally, for marginal perturbations with $\th=0$ one needs to go beyond the linear expansion.
We thus recover a very similar picture to the standard perturbative framework, but with a finite number of free couplings parametrizing the deviation from a NGFP.

%------------------------------------------------------------------------------
{\it Acknowledgments} -- I would like to thank A. Codello, D. Litim and D. Oriti for useful comments and corrections to the original manuscript.

%---------------------------------------------

%\bibliographystyle{JHEP-3}
%\bibliography{frge}
%---------------------------------------------

\providecommand{\href}[2]{#2}\begingroup %\raggedright
\endgroup

%------------------------------------------------------------------------------
\end{document}